\documentstyle[aps,preprint,epsfig]{revtex}
\begin{document}
\title{Cosmic dust grains strike again}
\author{Luis A. Anchordoqui}
\address{Department of Physics, Northeastern University, Boston,
Massachusetts 02115}

\maketitle
\begin{abstract}

A detailed simulation of air showers produced by dust
grains has been performed by means of the {\sc aires} Monte Carlo
code with the aim of comparing with experimental data.
Our analysis indicates that  extensive dust grain air showers
must yet be regarded as highly speculative but they cannot be
completely ruled out.

\noindent {\it PACS number(s): 98.70.Sa, 98.70.-f}
\end{abstract}

\newpage

It has long been known that small solid
particles (or dust grains) may be accelerated effectively at strong
collisionless shock waves -- popularly in supernovae -- \cite{spitzer}.
In addition, it was suggested years later that, even upon
destruction, a substantial fraction of the super-thermal debris
(those created in the pre-shock region) might re-enter the shock
waves to undergo additional acceleration \cite{epstein}.
These ideas have been recently strengthened by astronomical
observations \cite{bw} and by the analysis of primitive meteorites \cite{t}.

Despite some still open questions concerning the timescale for grain
destruction in the source environment and, later on,
during the trip to Earth,
from time to time dust grains have been considered as possible
progenitors of giant air showers. The earliest significant
contribution we are aware of is due to Herlofson \cite{herlofson}.
He argued that relativistic dust grains could indeed produce
extensive air showers, provided that in the first
interaction each constituent nucleon contributes with
an energy above $10^{14}$ eV.
The field then lay fallow for eighteen years until Hayakawa \cite{hayakawa}
tried to explain the exceptional event reported by Suga et al. \cite{suga}.
Prompted by this proposal several plausible suggestions came out in the
1970s \cite{dasgupta}.

Unfortunately, the values of the depth of
shower maximum registered by Haverah Park and  Volcano Ranch
experiments do not provide
an exact picture of showers initiated by dust grains \cite{linsley}.
Nonetheless, in view of the low statistics at the end of the spectrum
and the wide variety of uncertainties in these experiments, one may be
excused for reserving the judgment. In order to increase the statistics
significantly, the Southern Auger Observatory is  currently under
costruction: A surface array (that will record the lateral and temporal
distribution of shower particles) + an optical air fluorescence
detector (which will observe the air shower development in the
atmosphere) \cite{auger}. A major advantage of the optical device is
precisely its capability of measuring the depth for the maximum shower
development.

Whatever the source(s) of the highest energy cosmic rays,
the end of the spectrum remains unexplained by a unique consistent
model \cite{ehecr}.
This may be due to the present lack of precision of our knowledge of these
rare events, or perhaps, may imply that the origin and nature of ultrahigh
energy cosmic rays have more than one explanation. If one assumes the
latter hypothesis, relativistic specks of dust are likely to generate some
of the events.

The above considerations  have motivated us to re-examine the effects of
giant dust grain air showers. Before proceeding to discuss in detail how
these extensive showers evolve, it is useful to review some
key characteristics
of the mechanisms that could lead to the destruction of dust grains.
Notice that the opacity of the interstellar medium will impose lower
and upper limits on the value of the Lorentz factor $\gamma$ for the
primaries impacting on the Earth atmosphere.

An early investigation by Berezinsky and Prilutsky indicates that
the grains turn out to be unstable with respect to the development
of a fracture \cite{berezinsky}. On the one hand,
subrelativistic dust grains disintegrate in collissions with heavy
nuclei of the interstellar gas. On the other hand,
electrical stress induced by the photoelectric effect in the
light field of the Sun results in the mechanical disruption
and subdivision of relativistic grains. Doubts have also been
expressed about the prospects of surviving against heating arising from
photoionization within the solar radiation field \cite{mcbreen}.
The evaporation of the surface atoms \cite{elenskii-suvorov} and even the
capture of electrons from the interstellar medium have been suggested
as possible ways to reduce the accumulation of charges.

All in all, the path length up to the first break-up in favor of
these figures,
($\log \gamma \approx 2$ and initial radii between 300 - 600 \AA)
turns out to be of a few parsecs  \cite{berezinsky-book}, i.e., much less
than the characteristic size of the Milky Way. This entails that only
RX J0852.0-4622, the closet young supernova remnant to Earth
(distance $\approx 200$ pc) \cite{crb}, could be considered scarcely
far away.

Let us now turn to the discussion of dust grain air
showers (DGASs). Relativistic dust grains encountering the atmosphere will
produce a composite nuclear cascade. Strictly speaking, each grain
evaporates at an altitude of about 100 km and forms a shower of
nuclei which in turn produces many small showers spreading over a
radius of several tens of meters, whose superposition is observed
as an extensive air shower.
For  $\gamma \gg 1$, the internal forces between the atoms will be negligible.
What is more, the nucleons in each incident nucleus will interact almost
independently. Consequently, a shower
produced by a dust grain containing $n$ nucleons may be modelled by the
collection of $n$ nucleon showers, each with $1/n^{\rm th}$ of the
grain energy. Thus, recalling that muon production in proton showers
increases with energy
as $E^{0.85}$ \cite{gaisser}, the total number of muons produced by the
superposition of $n$ individual nucleons showers is,
$N_\mu^{\rm DG} \propto n (E/n)^{0.85}$, or,
comparing to proton showers, $N_\mu^{\rm DG} = n^{0.15} N_\mu^p$.
Of course, these estimates are approximations, but without
cumbersome numerical calculation one could naively select the event recorded
at the Yakutsk array (May 7, 1989) as the best giant DGAS candidate in the
whole cosmic ray data sample \cite{yakutsk,yakutsk2}.

In order to go a step further and test these qualitative
considerations, we have performed several
atmospheric cascade development simulations by means of the {\sc aires}
Monte Carlo code \cite{sergio}. Several sets of showers were generated,
each one for different specks of metallic nature, i.e.
with different Loretz factors. The sample was distributed in the energy
range of $10^{18}$ eV up to $10^{20}$ eV and was equally
spread in the interval of 0$^{\circ}$ to 60$^{\circ}$ zenith angle at
the top of the atmosphere.
All shower particles with
energies above the following thresholds were
tracked: 750 keV for gammas, 900 keV for electrons and positrons, 10
MeV for muons, 60 MeV for mesons and 120 MeV for nucleons and nuclei.
The particles were injected at the top of the atmosphere (100
km.a.s.l), and the surface detector array
was put beneath different atmospheric densities selected from the altitude of
cosmic ray observatories (see Table I).
{\sc sibyll} routines \cite{sibyll} were used to
generate hadronic interactions above 200 GeV. Notice that while
around 14 TeV c.m. energies the kinematics and particle production
of minijets might need further attention, for DGASs the
energy of the first interaction is reduced to levels where the
algorithms of {\sc sibyll} accurately match experimental data \cite{prdhi}.
Results of these simulations were processed with the help of {\sc aires}
analysis program. Secondary particles of different types and all
charged particles in individual showers were sorted according to
their distance $R$ from the shower axis.

In an attempt to examine our qualitative considerations
we first restrict the attention to the highest event
reported by the group at Yakutsk. Next, details of the most relevant
observables of the showers are given from the analysis of both the particles
at ground and those generated during the evolution of the cascade.

The Yakutsk experiment determines the shower energy by interpolating and/or
extrapolating the measurements of $\rho_{600}$
(charged particle density at 600 m from the shower core), a
single quantity which is known from the shower simulations to
correlate well with the total energy for all primary particle types.
In the case of the event detected on May 1989, the trigger of 50
ground-based scintillation detectors at 200 - 2000 m
from the shower core allowed to estimate a reliable value of
$\rho_{600} = 54$ m$^{-2}$ and a declination axis given by
cos $\theta$ = 0.52 \cite{yakutsk2}. Examining the lateral distribution
of our simulation sample, we
note that showers initiated by relativistic particles (log
$\gamma$ = 4 - 3.8) of energies in the range 36 - 38 EeV and with
orientation $\theta = 59^\circ$ are compatible with such a value of
$\rho_{600}$ \cite{antonov}. It is important to stress that the lower
and upper energy bounds are compatible with the event of interest
within 2 $\sigma$;
however, for these high Lorentz factors the interstellar medium is
extremely opaque to dust grains (again the reader is referred to
Ref. \cite{berezinsky-book}). In Fig. 1 we show the lateral
distributions of muons and charged particles. It is easily seen that the
predicted fluxes at ground level are partially consistent with those
detected at the giant array (see Fig. 2).

In what follows we shall discuss the main properties of DGASs.
The atmospheric depth $X_{\rm max}$ at which
the shower reaches its maximum number of secondary particles is the
standard observable to
describe the speed of the shower development. In view of the superposition
principle this quantity is practically independent of $n$, depending only
upon $\gamma$. For this reason the altitude of the $X_{\rm max}$ is generally used
to find the most likely mass for each primary.
The behavior of the $X_{\rm max}$ with increasing $\gamma$ in
vertical showers has been already discussed in detail
elsewhere \cite{linsley}. Here we shall extend Linsley's analyses
studying the dependence of the shower evolution with the incident angle.
In the last panel of Fig. 3 we show the numerical results
from  $10^{19}$ eV cascades initiated by nucleons with log $\gamma$ = 3
and different primary zenith angles. It can be observed that, at the same
total energy, an air shower induced by particles with an oblique incidence
develops faster than a vertical shower. Since muons are typically leading
particles in the cascade, the position
of $X_{\rm max}$ is also related to the relative portion of
muons in the shower. To illustrate this last point,
the resulting fluxes at ground level from the same events are shown in
first two panels of Fig. 3. As it is expected, the radial
distribution of the shower particles of inclined primaries
(mainly dominated by muons) is flatter than the distribution
of a vertical shower.
The opposite behavior points onto the supremacy of electrons and
positrons near the core. The density of these charged particles, however,
falls off rapidly with increasing core distance, dimming the electromagnetic
cascade.

To analyse the sensitivity of ground arrays to the shower parameters,
the radial variation of different groups of secondary particles (we
have considered separately $\gamma$, $e^+e^-$, and $\mu^+\mu^-$) was
studied at two
observation levels. In Fig. 4 we show the last steps of the evolution
of the lateral distribution along the longitudinal shower path.
It can be seen that there is practically no change in the
radial variation. Thus, we conclude that the flux of particles does not
have intrinsic sensitivity to the observation altitude until 1400 m.

Coming back to the general features of the longitudinal development,
the last exercise is to analyse the dependence of the atmospheric
shower profile with $\gamma$. To this
end we carry out  numerical simulations of vertical showers induced by
primaries with the same mass and different Lorentz factors, namely
log $\gamma$ = 2.5 to 3.2. The results are shown in the last panel of
Fig. 4. As expected, for the same primary mass the number of secondary
particles increases with rising $\gamma$, but it is interesting to note
that both cosmic ray cascades present similar shapes and peak around
the same atmospheric depth, i.e., $X_{\rm max} \sim$ 350
$\pm$ 47 g/cm$^{2}$  (consistent with the analysis of \cite{linsley}).

Putting all this together one can draw the following tentative conclusions:

\noindent (i) The Fly's Eye collaboration has presented evidence
indicating that typical extensive air showers above $\sim$ 1 EeV
develop at a rate which is consistent with a steep power law
spectrum of heavy nuclei that is overtaken at higher energies by a
flatter spectrum of protons \cite{flys-eye}. The group of the Akeno
Giant Air Shower Array  has reported 7 events above $10^{20}$ eV until
August 1998 \cite{takeda}. In general, the muon component agrees
with the expectation extrapolated from lower energies \cite{haya}.
However, the highest energy event recorded at Yakutsk 
($E = 1.1 \pm 0.4$ eV) seems to be a rare exception. It can be
readily noticed from Fig. 1 that the almost completely muonic
nature of particles detected may be explained by a 36 - 38 EeV
DGAS. Besides, if this speculation is true the lack of obvious
candidate sources at close distances leaves the origin of this
event still as a mystery.

\noindent (ii) The dependence of the DGAS longitudinal profile on $\gamma$
in the studied range is rather weak. The shower disc
becomes slightly flatter and thicker with decreasing primary energy and/or
increasing zenith angle. This is mainly due to the rising
content of muons among all charged particles in the inclined shower.

Dust is a very widespread component of diffuse matter in the Galaxy
with apparently active participation in charge particle acceleration.
If such acceleration is possible, dust grains may play an important
role in determining the composition of galactic cosmic rays.
The question of whether the opacity of the
interstellar medium could prevent relativistic dust grains from
reaching the Earth is as yet undecided.
Observation of giant DGASs would give an experimental and definitive answer
to this question. We recommend
that Auger search data be analysed for evidence of DGASs.

\hfill

I am grateful to D. G\'omez Dumm for helpful comments on the
original manuscript and to F. Zyserman for computational aid.  I
am also indebted to Yakutsk Collaboration, and particularly to M.
I. Pravdin for permission to reproduce Fig. 2. This work has
been partially supported by CONICET. The author is on leave of
absence from UNLP.

\begin{table}
\caption{Sites of giant air shower arrays}
\begin{tabular}{llll}
Experiment & longitude & latitude & altitude\\
\hline
Volcano Ranch & 35$^\circ$09'N & 106$^\circ$47'W & 834 g/cm$^2$\\
SUGAR & 30$^\circ$32'S & 149$^\circ$43'E & 1020 g/cm$^2$\\
Haverah Park & 53$^\circ$58'N & 1$^\circ$38'W & 1020 g/cm$^2$\\
Yakutsk & 61$^\circ$42'N & 129$^\circ$24'E & 1020 g/cm$^2$\\
Fly's Eye & 40$^\circ$'N & 113$^\circ$'W & 860 g/cm$^2$\\
AGASA & 38$^\circ$47'N & 138$^\circ$30'E & 920 g/cm$^2$\\
Auger & 35$^\circ$12'S & 69$^\circ$12'W & 875 g/cm$^2$
\end{tabular}
\end{table}

\begin{figure}
\label{yak}
\begin{center}
\epsfig{file=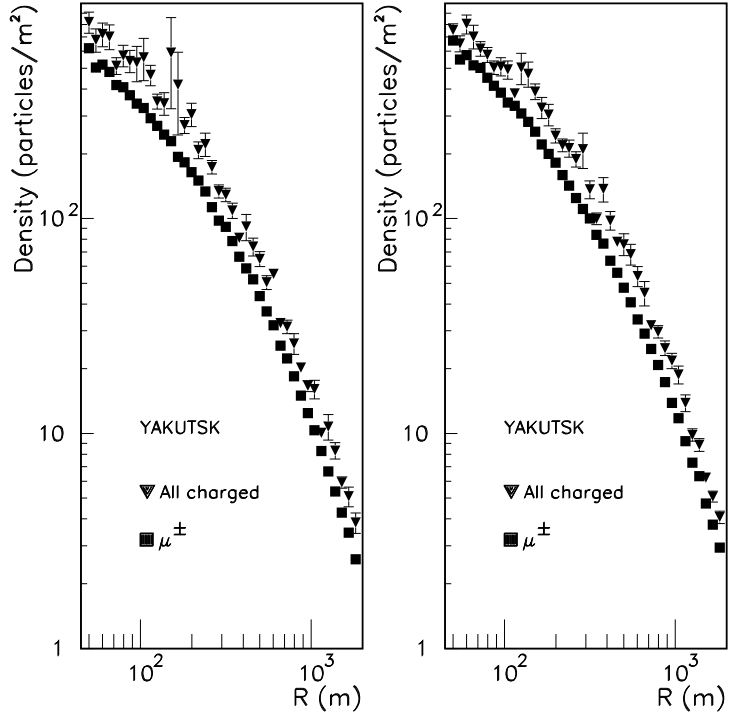,width=11.cm,clip=} \caption{From left to
right, (a) Lateral distributions of charged particles and muons
from {\sc aires} simulations of a $36$ EeV,  log $\gamma = 4$
DGAS. The error bars indicate the RMS fluctuation of the means.
(b) Id. with $ E= 38$ EeV and log $\gamma$ = 3.8.}
\end{center}
\end{figure}
\newpage

\begin{figure}
\label{yaky}
\begin{center}
\epsfig{file=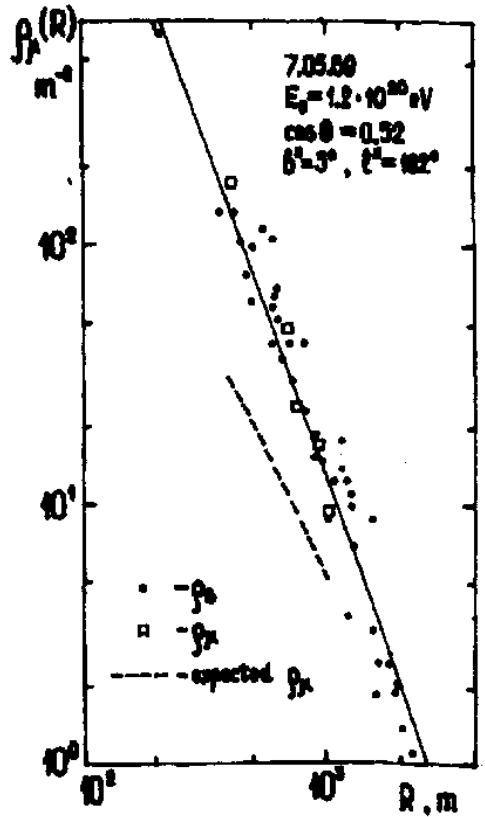,width=6.5cm,clip=} 
\caption{Density of muons recorded at Yakutsk by ground and 
underground detectors on May 7, 1989 at 13 h 23 min Greenwhich time. 
There is a remarkable contradiction with an extrapolation from results 
of the low energy region (dashed line). This figure was originally 
published in Ref. [18].}
\end{center}
\end{figure}
\newpage

\begin{figure}
\label{hp}
\begin{center}
\epsfig{file=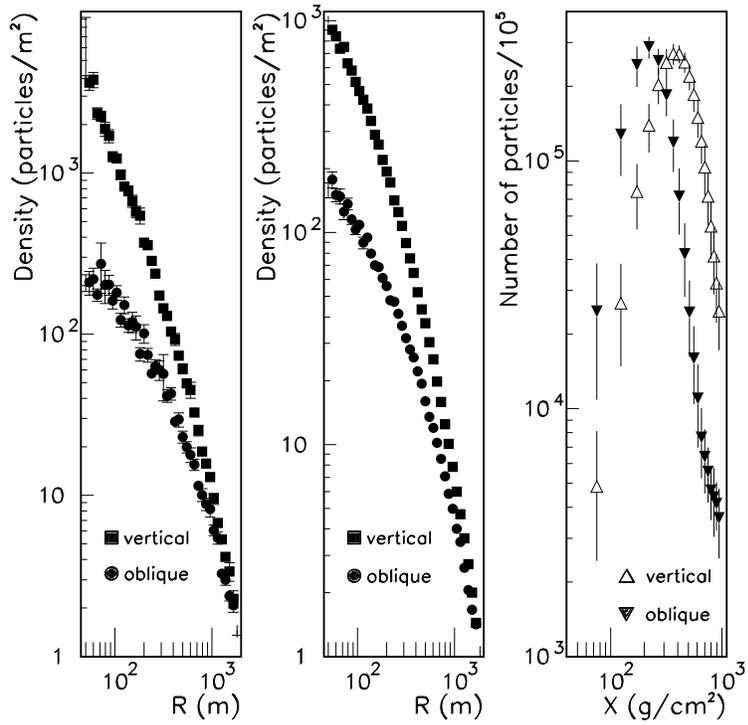,width=11.cm,clip=} \caption{From left to
right: (a) Lateral distribution of charged particles of DGASs of
10 EeV. The primaries with log $\gamma =3$ were injected
vertically and with 60 degree zenith angle. The figure also shows
the fluctuations measured in terms of RMS. (b) Idem for the case
of muons. (c) Longitudinal evolution of the shower profiles. The
error bars indicate the standard fluctuations.}
\end{center}
\end{figure}

\newpage

\begin{figure}
\label{auger-agasa}
\begin{center}
\epsfig{file=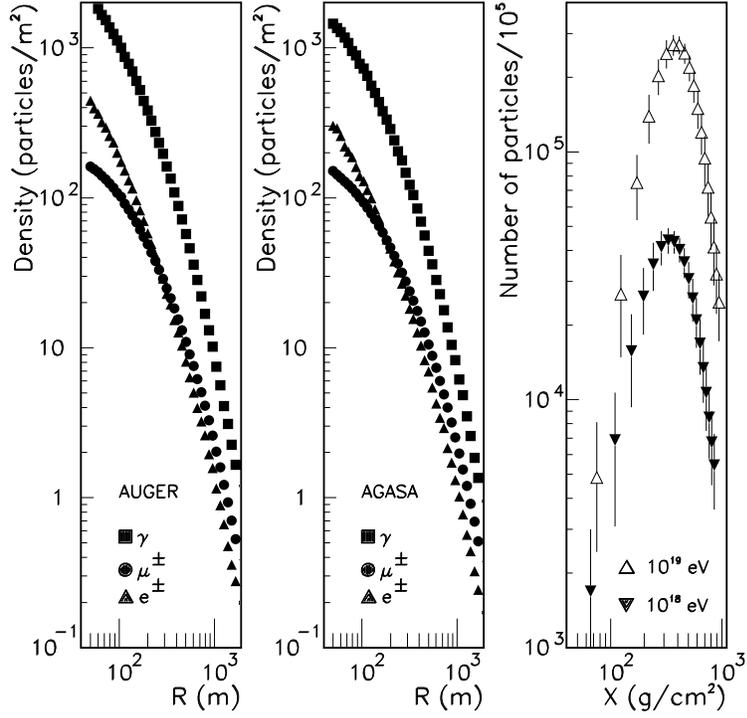,width=11.cm,clip=} \caption{Lateral
distributions of electrons, muons and gammas (at ground level)
from {\sc aires} simulations of $10^{18}$ eV vertical showers
($\gamma = 300$). The figure also shows the longitudinal cascade
development profile of $10^{18}$ - $10^{19}$ eV showers as would
be seen by the Auger fluorescence detector. The logarithm of the
Lorentz factors are 2.5 and 3.2 respectively. The error bars
indicate the standard fluctuations (the RMS fluctuations of the
means are always smaller than the symbols).}
\end{center}
\end{figure}
\end{document}